\documentclass[fleqn,10pt]{olplainarticle}

\title{On Controlled Change: Generative AI’s Impact on Professional Authority in Journalism}

\author[1, 5]{Tomás Dodds}
\author[2, 6]{Wang Ngai Yeung}
\author[3]{Claudia Mellado}
\author[4]{Mathias-Felipe de Lima-Santos}
\affil[1]{University of Wisconsin-Madison}
\affil[2]{Northeastern University London}
\affil[3]{Pontificia Universidad Católica de Valparaiso}
\affil[4]{Macquarie University}
\affil[5]{Leiden University}
\affil[6]{University of Oxford}


\begin{abstract}
Using (generative) artificial intelligence tools and systems in
journalism is expected to increase journalists' production rates,
transform newsrooms' economic models, and further personalize the
audience's news consumption practices. Since its release in 2022,
OpenAI's ChatGPT and other large language models have raised the alarms
inside news organizations, not only for bringing new challenges to news
reporting and fact-checking but also for what these technologies would
mean for journalists' professional authority in journalism. This paper
examines how journalists in Dutch media manage the integration of AI
technologies into their daily routines. Drawing from 13 interviews with
editors, journalists, and innovation managers in different news outlets
and media companies, we propose the concept of \emph{controlled change}
as a heuristic to explain how journalists are proactively setting
guidelines, experimenting with AI tools, and identifying their
limitations and capabilities. Using professional authority as a
theoretical framework, we argue that journalists anticipate and
integrate AI technologies in a supervised manner and identify three
primary mechanisms through which journalists manage this integration:
(1) developing adaptive guidelines that align AI use with ethical codes,
(2) experimenting with AI technologies to determine their necessity and
fit, and (3) critically assessing the capabilities and limitations of AI
systems.
\end{abstract}

\begin{document}

\flushbottom
\maketitle
\thispagestyle{empty}

\section*{Introduction}

Despite the recent hype and boom around artificial intelligence (AI),
particularly generative AI (GenAI), and the widespread speculation about
how these technologies could entirely transform the journalistic field \citep{Marr2024How}, the use of GenAI in the newsroom is not entirely new. In
fact, it builds on computational technologies that have been part of the
growing trend of automated journalism for at least a decade \citep{Pena2023Without, Yeung2024Automated}. For example, around the late 1990s, automatic transcription systems relying on speech recognition were already being deployed for English, French, German,
Spanish, and Japanese broadcast content \citep{Nouza2004Very}, and such
tools have only gotten more accurate over time. For illustration, the
word error rates for French have since decreased from 27.1\% with LIMSI
1998 Hub-4E Transcription System \citep{Gauvain1997LIMSI} to 9\% with
Microsoft Azure's STT \citep{Xu2021Benchmarking}. This already shows that GenAI
is not necessarily something new in journalism, but rather, the systems
behind these technologies are becoming increasingly more reliable and
persistent.

Though with varying terminologies (e.g., robot journalism or automated
journalism), these technologies generally refer to ``the process of
using software or algorithms to automatically generate news stories
without human intervention'' (\citealt{Graefe2016Guide}, p. 14). Similarly,
algorithmic journalism has also been described as the process whereby
increasingly more news content is ``generated by computer algorithms''
(\citealt{Shin2020User}, p. 1) or to ``assist and accelerate particular procedural
tasks, such as content creation, fact-checking, data processing, image
generation, speech conversion and translation, reducing the burden on
human and increasing efficiency'' (\citealt{Shi2024How}, p. 582). Common to
all of these definitions is that an increase in \emph{generation} is
both a key component and the end goal of robot journalism \citep{Lin2022One}. That is, do more, faster, with fewer humans involved.

In this article, we understand GenAI as an assemblage of machine
learning model-based tools, techniques, and systems that ``take what it
has learned from the examples it's been shown and create something
entirely new based on that information'' \citep{Carle2023Ask}. As such, these
technologies are capable of creating text, images, audio, and video
content to embellish news stories with only a short prompt required.
Yet, while \emph{generation} is not limited to texts, the crux of robot
journalism has remained mainly in Natural Language Processing (NLP) for
reducing human efforts spent on redundant, repetitive tasks and freeing
room for, e.g., human-driven and quality investigative journalism \citep{Stray2021Making}.

As early as 2008, the vision of technologizing newsrooms has already
been animated in higher education institutions and legacy media outlets.
For example, Columbia University and the \emph{New York Times} were both
at the forefront of this up-and-coming interdisciplinary advancement of
journalism with their joint degrees in computer science and journalism
and the Interactive News Technology (INT) desk, respectively (Anderson,
2013). Nevertheless, despite the significance of generation in such a
trend in the newsrooms, not until the explosive popularization of
Chat-GPT in late 2022 and its successor GPT-4 in 2023 academic studies
have explicitly referred to GenAI in journalism and its implication in
news production.

Since then, there has been a clear division between voices that seem
to hype the transformative power of (generative) artificial intelligence
in journalism and those who are more critical of the possible
implications of these technologies in newswork. For example, academics
have claimed that generative AI in journalism could usher in an era of
``potential transformation of journalism and media content,'' (\citealt{Pavlik2023Collaborating}, p. 84) or as \cite{Cherepakhin2024Responsible} puts it, ``AI will fundamentally
transform journalism and influence every stage of the process'' (p. 25). Yet the response from the industry itself has been more
moderated than that academic hyperbole. German journalist Anika Zuschke notes how, despite AI's ability to defeat a chess master in a few
seconds, the technology still struggles to recognize the importance and
ethics surrounding journalistic quotes \citep{Zuschke2024Citation}. As \cite{White2024AI} summarizes it,
the problem with AI systems today is that ``they do a poor job of much
of what people try to do with them, they can't do the things their
creators claim they one day might, and many of the things they are well
suited to do may not be altogether that beneficial.''

Nevertheless, even when these limitations are acknowledged, the rise
of GenAI has threatened professional authority in journalism \citep{Dodds2024Impact, Mellado2024Inteligencia}. Previous research on digital
technologies has already argued that journalistic authority is no longer
solely grounded in expertise or institutional backing but must now be
continuously renegotiated in the face of external challenges, such as
technological advancements and public skepticism \citep{Anderson2008Journalism, Robinson2007Someone}. As journalism evolves, so too must the strategies by
which authority is asserted and maintained. Yet, as AI-generated content
becomes more sophisticated, it challenges the foundations of
journalistic expertise by enabling non-professionals to create news-like
materials at unprecedented speed and scale. Journalists must not only
defend their authority against other actors in the digital ecosystem but
also against the encroachment of the technologies they are increasingly
expected to integrate into their work \citep{Dodds2025Knowledge, Simon2022Uneasy}

Hence, this article draws from interviews with 13 media workers and
journalists in the Netherlands and answers the following research
question: How are journalists navigating the integration of AI
technologies in newsrooms, and what role does professional authority
play in their approach to controlled change?

This article presents a historical overview of the development of
generative AI in journalism, identifying the key industry players and
significant technological developments in the field. We then present the
sample of our study and the analysis of the interviews with media
workers in the Netherlands. Our results show that journalists actively
engage in a state of ``preparedness'' to integrate AI systems into their
work process. This integration is characterized by a controlled
approach, where journalists anticipate and supervise the use of AI to
maintain their professional authority. Our research identifies three
primary mechanisms through which journalists manage this integration:
(1) developing adaptive guidelines that align AI use with ethical codes,
(2) experimenting with AI technologies to determine their necessity and
fit, and (3) critically assessing the capabilities and limitations of AI
systems. The study also explores whether this controlled change stems
from a perceived threat of losing agency, foresight about the rise of
AI, or a sense of empowerment in managing AI integration.

\subsection*{The Industry of Generative AI in Journalism}

Generative AI can produce human-like texts that are ``more human than
human'' \citep{Jakesch2023Human}, with linguistic features such as
first-person pronouns and the use of contractions. Yet, the results seem
to be less than successful for journalistic work. According to \cite{Longoni2022News}, for example, news readers are more likely to rate
AI-generated news headlines as more inaccurate and falser regardless of
their truthfulness when compared to a human-written headline.

The audience's dislike and distrust of AI-generated content (see also
\citealt{Mellado2024Inteligencia}) does not seem to deter the companies managing
newsrooms. Instead, these organizations continue to integrate AI into
their operations despite public skepticism. Of course, speech-to-text
technologies have also been recognized by tech start-ups as useful
products to sell to newsrooms with the promise of minimizing
journalists' labor for manual transcription. Not only does the
reliability and language flexibility, in terms of the variety of
languages and the ability of code-switching detection \citep{Ylmaz2018Code} of these systems increased, but the service providers have also
escaped from the academic bubble and slowly picked up by the large
platforms such as Google's Cloud Speech-To-Text, IBM's Watson Speech to
Text and Amazon's Transcribe.

This trend is not unique to transcription services and can be noticed
in other applications of generative AI in journalism. For example, in
the same year when Microsoft Azure was launched, Narrative Science Inc.
filed its first patent application for a narrative story generation
technology with Natural Language Generation or NLG \citep{Paley2020Applied}.
While Narrative Science has long transformed into a tech firm focusing
on narrative-based storytelling and data visualizations for businesses
and is now acquired by Salesforce Inc's Tableau, it was among the first
to inform journalists how NLG can revolutionize journalism with
automatic news text generation. Now, NLG is no longer a hidden gem in
journalism. Supported by Google's Digital News Initiative and
collaborating with various German research institutes, textOmatic is a
real-time text generation company that originated in Germany. With its
own SaaS-based text generator, textOmatic*Composer \citep{Bense2018Generating},
users are able to produce texts with standardized templates. Such
technology can optimize production processes by automating the curation
and cleaning of data for news with repetitive structures \citep{Kotenidis2021Algorithmic}, with economic and sports news being two working examples
\citep{Kanerva2019Template}.

On the other hand, regarding social networking sites (SNS),
@wpolympcsbot is perhaps a prototypical demonstration of how generative
AI can be merged within social media ecology. During the 2016 Olympics,
taking advantage of data extraction techniques and template-based text
generation, \emph{The Washington Post} implemented a real-time sports
reporting bot on Twitter (now X). When commenting on the role of AI in
journalism, the outlet's digital project department claimed that the
``AI will be focusing on simple aspects of the games such as scores and
medal counts, mostly data-related facts, thus freeing up the human
reporters to tackle more complex stories'' \citep{Caughill2016Washington}. A few years
later, \emph{The Guardian Australia} published its first automatically
generated economic news with its in-house reporting system, ReporterMate \citep{Schapals2020Assistance}. Focusing on political donations, the
article is accompanied by drop-head and in-text data visualizations
summarizing the most prominent donors in Australian politics, both of
which are automatically created. Though ReporterMate was short-lived,
and the language is relatively simple and coarse-grained given its
experimental nature, such full automation was ahead of its time and
sparked worldwide attention when it was released. Although whether
generative AI is still merely ``data related'' is up to debate,
particularly with the rise of Large Language Models (LLMs) such as GPT-4
and Gemini, narrative generation technologies are unquestionably more
competitive than they were just half a decade ago.

\subsubsection*{Maximizing and experimenting with audience engagement: Newsbots
and chatbot news}

In Briggs' \emph{Journalism 2.0} \citep{Briggs2007Journalism}, the first chapter teaches
students how to use Really Simple Syndicator (RSS) feeds to
automatically collect information from a source website (e.g., a news
site) in XML file format. This marks the cornerstone of digital
journalism and robot journalism, as information does not have to be
actively sought after and aggregated as long as the target website has
the RSS in place. Notwithstanding that RSS is not a bot in itself, the
underlying conceptual mechanism of data fetching-parsing is analogous to
a real-time data scraping bot. Newsbots are now more sophisticated than
a scraper; they can also offer opinions, entertain users, and
investigate information \citep{Lokot2016News}. Enabled by
technical advancements within NLG, newsbots can now take over recurring
and often dull tasks like sports reporting, as previously illustrated by
@wpolympicsbot, with an infusion of nearly indistinguishable language
from humans.

Remarkably, what one should also pay attention to is how chatbots and
conversational bots are integrated into contemporary journalism (see
\citealt{Resendez2023Hey}). With social media platforms becoming the
\emph{de facto} element of journalism, newsrooms have also wittingly
adapted to this change by distributing news on Facebook, Twitter, and
other environments alike. For example, \emph{The New York Times}
collaborated with Smart Politics, an NGO aiming at tackling polarization
in the U.S., to develop a game-like chatbot called Angry Uncle Bot \citep{Zarouali2021Overcoming}. Strictly speaking, its algorithm is a
straightforward IF-THEN operation. Nonetheless, NLG models (e.g.,
GPT-3.5-Turbo) can be implemented to generate contextual responses to
increase the perceived realism of the bot.

\subsubsection*{Automating content tagging, title suggestions, keyword generation
and automated analytics}

The premise of Google's search engine, if not all for-profit search
engines, is to rank webpages by their quality and efficiently present
the best ones to the users \citep{Brin1998anatomy}. The ranking algorithm
is based on a calculation of an array of quality indicators, and its
resulting score, the so-called PageRank, is approximated with features
such as backlinks, anchor texts, and HTML content of a given webpage.
The process of altering web content for a better ranking score is now
globally known as Search Engine Optimization (SEO), and the quality
indicators now also include page speed, freshness, and other technical
elements. To compete with other news outlets in the digital world,
journalists ought to familiarize themselves with SEO. As research
engines' ranking algorithms complicate over time, SEO requires immense
technical expertise and trend awareness to be successful. Generative AI
is, unsurprisingly, instrumental in this regard. A study investigating
the news ecosystem in the United Arab Emirates by \cite{Ahmad2023News}
has elucidated the capacity of generative AI to assist journalists in
choosing words `wisely' to strengthen the outlet's digital presence.
Techniques such as article keyword generation, automated analytics, and
alt-text generation \citep{Chiarella2020Using} allow journalists to
magnify engagement (e.g., click rates) and searchability on search
engines. Moreover, another study indicates how the combined force of
computer vision and text generation can help automate image and video
tagging \citep{Zhang2022CREATE}, which is available on some social media
platforms such as Twitter. Another traffic-generating element of online
news articles is the article titles themselves. Thus, news headline
generation \citep{Gu2020Generating, Li2022Faithfulness} is a desirable tool that
journalists would like to pursue. However, most newsrooms do not possess
the technicality to build an in-house tool to do so, and therefore they
often must seek help from external software developers and specialized
tech start-ups, which are dominantly proprietary (e.g., Hypotenuse AI)
or freemium (e.g., Sassbook), and seldom free (e.g., the Hoth). Hence,
the question arises as to how this growing dependency might impact
professional authority within media organizations.

\subsection*{Contemporary challenges to professional authority}

Professional authority in journalism refers to the recognized power and
influence that journalists hold within their field, shaped by their
expertise, adherence to ethical standards, and their gatekeeping role in
disseminating information \citep{Carlson2017Journalistic, Ornebring2013Anything, Shoemaker2013Mediating}. Historically, this authority has been rooted in the trust
journalists build with their audiences by providing accurate, impartial,
and timely information \citep{Weaver2016Changes}, which positions them
as guardians of public knowledge and mediators between society and
different institutions \citep{Mellado2020Beyond, Waisbord2018Truth}.

Despite the foundational principles that support journalistic authority,
contemporary challenges continue to emerge. The proliferation of ``fake
news'' and misinformation has directly undermined public trust in
journalistic institutions \citep{Christofoletti2024Trust, Dodds2021Structures}. Audiences
are now more critical of where and how they receive information \citep{Meijer2013Valuable}, and journalists must compete in an environment where alternative
narratives---often unverified---spread rapidly across digital platforms \citep{Ross2022Its}. In the context of generative AI, this
authority faces unique challenges as AI begins to take on roles
traditionally held by human journalists, particularly in content
creation, data processing, and reporting.

Generative AI systems, like those built upon Natural Language Processing
(NLP) and Large Language Models (LLMs), can now produce texts, images,
and even audio that mimic human creations, raising questions about the
future of journalistic labor and the professional authority journalists
hold in their craft, where the boundaries between human journalism and
machine-generated content begin to blur \citep{Yeung2024Automated}.

In this context, the introduction of generative AI into newsrooms has
caused concern among journalists, many of whom fear that their authority
could be diminished as AI technologies take on increasingly complex
tasks \citep{Gutierrez2023AI, Simon2022Uneasy}. These fears are not
unfounded. AI systems can quickly generate news stories from structured
data, summarize complex reports, and even generate headlines that are
optimized for digital engagement. Yet, as our findings suggest,
professional authority in journalism is not necessarily at risk of being
eroded by AI. Rather, it is adapting to the new media landscape where
journalists act as supervisors, gatekeepers, and critical interpreters
of AI-generated content.

In Dutch newsrooms, as highlighted by this study, journalists are
developing adaptive guidelines for integrating AI into their work. These
guidelines evolve as the technology advances, ensuring that AI is used
responsibly and ethically. Journalists in these newsrooms view AI as a
tool to augment their work, not replace it. By maintaining control over
the AI tools they use, journalists preserve their authority, positioning
themselves as essential arbiters of the news production process.

The concept of ``controlled change'' is then relevant in the discussion
of AI's integration into journalism in which journalists are not
passively adopting AI but are actively managing and supervising its use,
ensuring that AI complements rather than replaces their roles \citep{Fridman2023How}, testing its capabilities while remaining mindful of its
limitations. In some cases, journalists explicitly label AI-generated
content, indicating that a human journalist has supervised or edited the
output \citep{Zier2024This}. Labels such as ``written by a human'' or ``edited
by a journalist'' are becoming more common, serving as markers of
professional authority in an increasingly automated field \citep{Waddell2019Attribution}.

\subsection*{Professional authority and journalism}

Theories of professional authority, particularly those stemming from Max
Weber, emphasize that authority derives from a combination of expertise,
social legitimacy, and institutional norms. Journalism, as a profession,
has traditionally held authority by being seen as the ``fourth
estate''---a vital check on power and an essential component of
democratic society \citep{Sampedro2024Imaginary}. Journalists were
gatekeepers, ensuring that the information reaching the public was
vetted and verified \citep{Shoemaker2013Mediating}. Under this umbrella,
journalists are recognized as professionals not only because they
possess technical skills but also because their work carries significant
social consequences \citep{Vos2019Discursive}.

However, as generative AI increasingly enters newsrooms, all professions
face a redefinition of authority. Outside the field, theories of
automation \citep{Coombs2020strategic, Zirar2023Worker} suggest that as
tasks become automated, professionals must shift their focus to areas
where human expertise is indispensable---particularly in ethical
judgment, nuanced interpretation, and creative problem-solving. Emerging
theoretical perspectives on automation \citep{Brynjolfsson2014Second}
also suggest that professionals will increasingly work alongside
machines, providing oversight and critical input.

Within the field of journalism, \citealt{Carlson2017Journalistic} introduced a relational
model according to which professional authority is no longer solely
determined by journalists themselves but is now contested through
interactions with audiences, sources, and technologies. Carlson argues
that journalistic authority is not a fixed property but is constantly
negotiated based on the context in which journalism operates. Under this
umbrella, he asserts that journalists must legitimize their role amidst
competing voices, such as social media influencers, bloggers, and
algorithmically driven platforms. Authority is thus not simply about the
possession of expertise but the ongoing processes that journalists
engage in to remain relevant and trusted in the digital age.

Along with Carlson, scholars such as \citealt{Anderson2008Journalism} and \citealt{Robinson2007Someone}
investigate the implications of digitalization for the authority and
autonomy of journalists in environments where news production is
becoming more automated and data-driven and how transparency and
audience engagement practices can reinforce or challenge journalistic
authority, particularly in an era of skepticism towards traditional
media.

\citealt{Anderson2008Journalism}, for example, introduces the idea that journalistic
authority is increasingly tied to institutional adaptability---how well
a newsroom can integrate new technologies and address the changing needs
of audiences while maintaining high professional standards. His work
shows that institutional resilience plays a critical role in maintaining
journalistic authority. He argues that while individual journalists hold
authority, news organizations as institutions also play a pivotal role
in shaping that authority, maintaining editorial control while adopting
new technologies. In other words, while automation and algorithmic
curation can aid in news production, the authority of journalism is
ultimately preserved through human editorial oversight.

\citealt{Robinson2007Someone} also provides insights into how journalists can
reinforce their authority through audience engagement and transparency.
According to Robinson, by making their editorial processes
visible---such as explaining how stories are sourced, verified, and
reported---journalists can strengthen their professional authority.
Journalists' ethical responsibility, therefore, becomes a foundation
upon which their authority rests.

Overall, journalism scholars highlight that journalistic authority is no
longer solely grounded in expertise or institutional backing but must
now be continuously renegotiated in the face of external challenges,
such as technological advancements and public skepticism. As journalism
evolves, so too must the strategies by which authority is asserted and
maintained. In journalism, this redefinition means that news
professionals must reposition themselves not as mere content creators
but as ethical supervisors of AI-generated content, where human values
must remain central in the integration of new technologies into
professional practices.

\section*{Methodology}

This study employs a qualitative research design to explore how Dutch
journalists are integrating generative AI tools into their workflows
while maintaining professional authority. Given the speed and nature of
AI adoption in Dutch journalism, we chose to conduct semi-structured
interviews as the primary data collection method. This approach allows
us to pursue in-depth conversations with participants while ensuring
interview comparability \citep{Brinkmann2016Methodological}.

We conducted 13 semi-structured interviews with professionals in various
key positions across Dutch news organizations, including multiple
journalists, foreign editors, editors-in-chief, deputy editors-in-chief,
strategy directors, innovation managers, and data chiefs.

Participants were recruited from the following organizations:
\emph{AD}, NPO, NOS, ANP, RTL, \emph{NRC}, \emph{Het Parool}, \emph{De
Volkskrant}, and DPG. To maintain confidentiality, we do not disclose
the specific positions associated with each organization in the text.
The interviews focused on participants' experiences with AI tools,
strategies for integrating these tools into daily routines, and the
guidelines established to govern their usage. Interviews were
transcribed and analyzed thematically by two research team members,
allowing us to identify recurring concepts and practices, especially
around the theme of professional authority. This sampling strategy aimed
to capture diverse perspectives within the Dutch media landscape,
following the principle of maximum variation sampling \citep{Suri2011Purposeful}.

The interviews were conducted between May and August 2023 with the
help of a research assistant via secure online video calls. On average,
the interviews lasted 34:22, with the longest lasting at 56:52 and the
shortest at 19:59. All the interviews were recorded with the
participant\textquotesingle s consent and declared in the consent form
each participant signed. All the interviews were transcribed verbatim
and translated from Dutch to English for analysis.

As a team, we developed an interview guide while reviewing the
existing literature on AI in journalism, focusing mainly on the
Netherlands and the professional and organizational challenges most
often associated with its implementation. Consequently, our questions
were designed to explore how media workers in the Netherlands perceive
and navigate AI systems being integrated into their routines and the
larger institutional factors influencing these processes (see Annex
1).

Afterward, the analysis followed a multi-step coding process inspired
by grounded theory \citep{Charmaz2006Constructing}. Initially, we conducted open coding
by systematically reading and labeling the interview transcripts,
ensuring that key themes emerged directly from the data. As a team, we
constantly discussed the categories as they appeared. Subsequently, we
applied axial coding to group the open codes into broader thematic
categories. Following our research question, we focused on how AI
adoption is framed within Dutch newsrooms. Finally, selective coding was
used to establish overarching themes that connect the identified
categories.{~ }Through this analysis, we developed the heuristic of
``controlled change'' to conceptualize journalists' deliberate and
supervised integration of AI in their work. At this stage, the
saturation point was reached, as no new themes emerged from additional
interviews \citep{Guest2020simple}.

\section*{Findings}

Our results show how journalists, editors, and managers within Dutch
newsrooms are testing, negotiating, and adapting AI-based technologies
into their routines and workflows. The findings are categorized into
three main areas: AI experimentation, adaptive guidelines, and
recognizing AI's capabilities and limitations. These areas reflect the
nuanced approaches that Dutch journalists are taking to manage the
challenges and opportunities presented by AI in their daily work.

\subsection*{Adaptive Guidelines}

As AI systems become increasingly integrated into journalistic
practices, Dutch newsrooms are developing and refining guidelines to
manage their use responsibly. However, far from static, during our
interviews, these guidelines were often described as ``living
documents'' that evolve in response to new developments and insights.
One editor explained, ``This entire protocol is a living
document\ldots{} you can always think differently about some things six
months from now'' (P8). Journalists and editors argued that this
adaptability is crucial, given how quickly GenAI is mutating and
advancing outside newsrooms and the ethical considerations necessary to
integrate these systems into journalism work.

According to some of our interviewees, creating these guidelines
involves extensive discussions within news organizations, not only among
editors and managers but also with input from digital media departments
and external experts. These conversations aim to delineate the ethical
and professional boundaries of AI use in journalism. As a tech
journalist said during our interview, ``I think we are looking for
boundaries to be set about what we all consider to be ethically
correct'' (P5). Highlighting the collaborative approach to defining what
is considered responsible AI usage, an editor (P8) recounted:

``What we have done a lot in the past two months is to bring in quite
a bit of knowledge from our own people to the news floor. We have a
department here where the digital products are made. Where data
scientists work with several people who know a lot about AI. So, we
brought them over to the news floor a bit more and asked them to help us
think about AI for a second. About what is possible. What are some
ethical and moral frameworks? Of course, we have to sketch them
ourselves. {[}\ldots{]} But what are the boundaries of what we do? What
we find exciting, what we don't find exciting at all. That's what we are
looking into. Having a living document just for here, internally, in
which we outline some preconditions and guidelines of what we consider
responsible journalistic use of AI.''

These discussions focus on the extent to which AI should be allowed to
handle tasks traditionally performed by journalists, like summarizing
complex investigative reports. ``You pop those 30 pages into a GenAI
model, and then you ask, summarize it for me. You leave the thinking,
which is the journalists' work, to a model, to AI. You skip a lot of
processes for yourself'' (P5) one journalist pointed out, stressing the
potential risks of over-reliance on AI for tasks that require critical
analysis and judgment. He later added, ``You risk blindly trusting such
systems. We know that those systems are absolutely not safe and that
many mistakes are made. So, would it be good to discuss with each other
whether it is desirable or not? I get paid for thinking and critically
reviewing documents.''

During our interviews, we found that guidelines, while still
evolving, aim to balance encouraging experimentation with AI and
maintaining the core principles of journalism. For example, our
interviewees pointed out that one of the most important elements of
these guidelines was ensuring that AI complements, rather than replaces,
the human elements of journalism. ``Everyone can play journalist.
{[}\ldots{]} So soon, some stories will explicitly add, `This story was
typed by a human,' so to speak. And in that, you should understand that
the writer has done journalistic work, namely that he has checked what
he writes and talked to people, real people, not AI people. And all
those things that are self-evident, you may have now to emphasize them''
(P1), a journalist asserted, underscoring the importance of preserving
the integrity and authenticity of journalistic output in the face of
technological advancements. When reflecting on creating guidelines, our
interviewees often highlighted that guidelines are guardrails that
encouraged them to test new tools while remaining vigilant about the
potential ethical implications.

\subsection*{AI Experimentation}

Our interviews revealed that journalists in Dutch newsrooms are actively
testing AI tools to enhance efficiency, particularly in routine tasks.
This testing, however, occurs in a controlled environment. As an
editor-in-chief put it, ``I don't see things changing enormously.
Rather, I see a controlled change occurring. We really aim to support
journalism with AI, not replace it'' (P3).

Part of this controlled change is gathering journalists and thinking
critically about how AI can serve journalists in their work. In one of
the newsrooms where we conducted interviews, we found journalists and
editors getting together to list what they wished AI could do for them:
``We made a wishlist, which we have submitted to our IT department. We
hope they can pay attention to this and start testing'' (P3).

As this quote reflects, experimentation inside newsrooms often takes
place collaboratively. The same editor also explained: ``We are doing
another test tonight, where we are going to write prompts together.
{[}\ldots{]} We have small working groups where we work together with
colleagues'' (P3).

One of the recurring themes in our interviews regarding AI
experimentation in the newsrooms is how surprised journalists were with
the speed AI brings to the table, allowing journalists to perform tasks
such as summarizing large amounts of information quickly. Regarding the
first tests for using AI tools for transcription, a journalist argued,
"It just works super-fast. And it's more accurate than doing it
manually, I think'' (P1). AI for generating summaries and quick recaps
has seemingly become standard practice, mainly when dealing with complex
and time-sensitive topics like the ongoing conflict in Ukraine. ``If I
need to do a quick recap of what happened last night in Ukraine,'' the
same journalists mentioned (P1), ``I can throw in links, and it's done
fast.'' However, we also found journalists aware of the possible
pitfalls of speed while conducting these tests, with one editor-in-chief
arguing: ``We want to be right, we want to be careful. Fast is second''
(P3).

AI is not only being tested for summarization but also as a
brainstorming partner. We found that journalists have begun to test AI
as a creative assistant, using it to generate ideas and explore new
angles for stories. For instance, one journalist described how they
asked AI for suggestions on articles related to the anniversary of the
war in Ukraine: ``I asked ChatGPT\ldots{} what articles can we write
about a year of war in Ukraine. And then it comes up with ten
suggestions'' (P1). While not all suggestions were helpful, the process
helped to refine ideas and identify potential storylines that might not
have been immediately obvious.

Despite these benefits, journalists argued that, after
experimentation, there was a clear understanding that AI is a tool
rather than a replacement for human creativity and journalistic
judgment. AI's role inside Dutch newsrooms is often seen as a ``stepping
stone'' or ``time saver,'' helping to manage the broader strokes of
content creation while leaving the nuanced, creative work to the
journalists themselves.

The interviews also revealed a cautious approach to AI-driven
experimentation. While AI can significantly speed up specific processes,
journalists remain aware of its limitations. As a data editor argued,
``It is the experts, the writers, who can distinguish themselves with
good journalism, using the right sources. I don't see ChatGPT suddenly
replacing that'' (P7). AI may suggest widely applicable ideas that could
save time in brainstorming sessions, but journalists recognize that
these ideas often lack the specificity and depth required for
high-quality journalism.

\subsection*{Capabilities and Limitations}

While AI offers numerous capabilities that can enhance journalistic
work, in our interviews, we found a strong awareness among Dutch
journalists of its limitations. One of the most significant concerns
highlighted by our interviewees is the potential erosion of the
journalistic process, particularly the verification and fact-checking
that have become central to the profession. ``You cannot let a computer
do your thinking, do the journalistic process of gathering news,
reading, trying to interpret, summarize, asking questions, looking for
things that are not correct, testing the truth'' (P5), one tech
journalist argued, noting that the use of AI should not compromise the
credibility of the work produced. The ability to verify information,
consult with human sources, and ensure that all aspects of a story are
accurate remains, at least in these newsrooms, a fundamental aspect of
journalism that AI cannot yet replicate.

However, journalists also recognized the potential of AI to assist in
different areas of content creation and analysis. During our interviews,
we found that AI tools can be particularly useful in tasks such as
rediscovering archives, analyzing past content, and even visualizing
complex ideas before they are fully developed. As one of the innovation
directors for a broadcast newsroom put it, ``You can use GenAI for your
entire process {[}\ldots{]}. You can now visualize ideas that you have
from the story much more easily. Write scripts, create screenplays,
invent scenes'' (P2). GenAI's ability to quickly prototype and test
ideas, including AB testing headlines, was seen by our interviewees as a
significant advantage. As a data editor also mentioned:

``If we can use {[}AI{]} as a tool to expand the AB testing of
headlines, from 15 articles that now we do manually to 30 that are done
automatically, I wouldn't mind. But I understand there is a fear that
the robot or the AI will take over things'' (P7).

However, we also found a cautionary note about the overuse of AI,
particularly in ways that may not align with the strategic objectives of
our interviewees' news organizations. ``You need to base your use of AI
on your objectives and strategy, not just on opportunities,'' (P2) one
of the innovation directors advised. Interestingly, the importance of
intentional and purposeful use of AI technologies was a recurrent theme
among our interviewees. This approach seems to ensure that AI is being
employed in ways that genuinely add value to the journalistic process
rather than simply adopting new tools for the sake of innovation.

The limitations of AI were also a recurring theme in our
conversations. While AI can assist with several tasks, it is not yet
capable of replacing the nuanced judgment and creativity that define
high-quality journalism. As one of the data editors put it:

``It is a broad playing field, and what we are looking at right now is
where AI can be a tool for editors and journalists, apart from the fact
that we also have daily discussions about AI and its applications in our
work. {[}\ldots{]} No one really knows where it's going. I don't think I
have spoken to anyone who now knows where it will be in three years from
now'' (P7).

Despite this uncertainty, there is a clear recognition that AI has a
role in the industry, particularly as a tool for enhancing productivity
and efficiency. However, its integration must be carefully managed to
avoid undermining the fundamental principles of journalism. This mixture
of optimism and caution was present throughout our interviews. The
motivations behind this mixture are best explained in this quote by a
tech journalist: ``I think it's very good that we are very critical of
AI because the moment you hand over your creativity to an AI system, it
is very difficult to go back to a situation where you didn't use it.''

\section*{Discussion}

While AI offers new possibilities for the future of journalism, Dutch
journalists are approaching its integration with a mixture of optimism
and caution. They are actively experimenting with AI, developing
adaptive guidelines, and critically assessing its capabilities and
limitations to ensure that its use enhances rather than detracts from
journalism's core values.

We draw from interviews with 13 media workers in the Netherlands to
answer our research question: How are journalists navigating the
integration of AI technologies in newsrooms, and what role does
professional authority play in their approach to controlled change?

In this article, we have argued that AI itself is not new to
newsrooms (see also \citealt{Diakopoulos2019Automating}). We have shown how automation,
algorithms, and other forms of computational assistance have been used
inside newsrooms for years, particularly in areas like data-driven
journalism, personalization, and audience engagement \citep{Pena2023Without, Yeung2024Automated}. The hype of recent generative AI
tools, however, resides in their capacity to produce text, images, and
other forms of content autonomously. However, Dutch media workers
demonstrated significant forbearance; rather than rushing to adopt these
tools, they exhibited restraint, experimenting carefully and
implementing AI in ways that are consistent with established
journalistic values.

Our results identified three mechanisms by which journalists managed
this integration. First, and in line with previous literature,
journalists have developed adaptive guidelines that align AI use with
existing ethical codes \citep{Santos2024Guiding}. During our
interviews, journalists and editors acknowledged the need for AI
governance, emphasizing transparency in AI-generated content and setting
boundaries on its application. These guidelines reinforce the idea that
AI tools serve as a complement to, rather than a replacement for, human
expertise.

The second mechanism identified in this research is a phase of
experimentation, where journalists are assessing AI technologies to
determine their suitability and usefulness inside the newsrooms. We
found journalists taking a hands-on approach to understanding AI's
capabilities and limitations, allowing them to identify practical uses
while maintaining a clear sense of what human oversight is required.

The third mechanism we identified from our interviews is a critical
assessment of AI systems. Journalists are not only interested in how AI
can improve efficiency or enhance storytelling but also focus on the
broader implications of integrating such systems. In line with previous
literature on this issue \citep{Ross2022Its, Simon2022Uneasy}, this
scrutiny helps prevent overreliance on AI and maintains the
human-centered judgment critical to journalism's mission.

In this article, we introduce the concept of \emph{controlled change}
as a heuristic to explain how newsrooms, at least in the Netherlands,
manage AI integration. Controlled change suggests that journalists are
not passively adopting new technologies but are actively setting the
terms under which AI will be used. As others have also argued in the
past \citep{Gutierrez2023AI}, this deliberate management of AI
integration illustrates a proactive stance where journalists take steps
to incorporate AI while protecting their professional autonomy and
authority.

While these findings support previous studies claiming that
journalists are cautiously approaching AI, controlled change extends
these discussions by emphasizing the structured, negotiated process of
AI adoption. Importantly, the three mechanisms we identified in this
article contrast with deterministic views of technological integration
that either portray AI as an inevitable disruptor or assume outright
rejection by journalists. Indeed, our results show that by actively
experimenting with AI while setting ethical and operational boundaries,
journalists are neither passively adopting new technologies nor
resisting them wholesale. This aligns with previous research on boundary
work in journalism \citep{Belair2018Boundary, Carlson2017Journalistic}, where
professional authority is continuously reasserted through negotiated
interactions with emerging technologies.

That is, the three mechanisms we have identified in this article
contribute to theoretical debates on professional authority by
illustrating how journalists maintain control over AI's role in news
production. The adaptive guidelines they develop assert editorial
oversight, experimentation allows them to shape AI's newsroom functions
in ways that align with professional values, and critical assessment
ensures that AI remains a tool rather than a determinant of journalistic
work. This structured approach differentiates controlled change from
more general discussions of technological adaptation in journalism,
offering a framework for understanding how news professionals actively
shape, rather than merely respond to, technological developments.

Finally, one of the most significant findings of this study is that
journalists do not see their professional authority as inherently
threatened by AI. Instead, they view the integration of AI as an
opportunity to redefine their roles in the newsroom. Acting as
supervisors, gatekeepers, and critical interpreters of AI-generated
content, journalists in the Netherlands are reconfiguring their
positions to maintain oversight over AI's influence on news production.
This evolution of roles shows that professional authority is adaptable
and resilient, as previously argued by \citealt{Robinson2007Someone} and \citealt{Carlson2017Journalistic}. When it comes to AI systems, professional authority in
journalism is shaped by ongoing negotiations rather than passive
acceptance of technological change.

\bibliography{ref}

\end{document}